# Ultra-low voltage resonant tunnelling diode electroabsorption modulator

J. M. L. FIGUEIREDO

Faculdade de Ciências e Tecnologia, Universidade do Algarve, Campus de Gambelas, 8000-117 FARO, Portugal

C. N. IRONSIDE, C. R. STANLEY

Department of Electronics and Electrical Engineering, University of Glasgow, Glasgow G12 8LT, UK

**Abstract.** Embedding a double barrier resonant tunnelling diode (RTD) in an unipolar InGaAlAs optical waveguide gives rise to a very low driving voltage electroabsorption modulator (EAM) at optical wavelengths around 1550 nm. The presence of the RTD within the waveguide core introduces high non-linearity and negative differential resistance in the current-voltage (I-V) characteristic of the waveguide. This makes the electric field distribution across the waveguide core strongly dependent on the bias voltage: when the current decreases from the peak to the valley there is an increase of the electric field across the depleted core. The electric field enhancement in the core-depleted layer causes the Franz-Keldysh absorption band-edge to red shift, which is responsible for the electroabsorption effect. High frequency ac signals as low as 100 mV can induce electric field high speed switching, producing substantial light modulation (up to 15 dB) at photon energies slightly lower than the waveguide core band-gap energy. The key difference between this device and conventional *p-i-n* EAMs is that the tunnelling characteristics of the RTD are employed to switch the electric field across the core depleted region; the RTD-EAM has in essence an integrated electronic amplifier and therefore requires considerably less switching power.

1. **Introduction**

High-speed and long haul optical fibre transmission systems require high-performance external optical modulators due their reduced wavelength chirping compared to high-speed direct modulation of diode lasers [1][2]. The external optical modulators are mainly based on III-V compound semiconductors and lithium niobate (for a revision see [3]). The former permits integration with laser diodes and photodetectors and has potential for improved stability, compactness, and cost [1][2].

The high precision growth techniques for semiconductor layers, in particular Molecular Beam Epitaxy (MBE), produce high precision growth allowing the fabrication of device with characteristics that are crucially dependent on layer thickness and tunnelling effects. Devices employing the tunnelling effect have been extensively studied and are now being considered for memory devices [4]. Furthermore, it has been demonstrated that III-V semiconductor tunnelling devices can be integrated with silicon CMOS technology, and that CMOS electronics can drive tunnelling devices [5]. The physics and progress in electronic applications of resonant tunnelling devices (RTDs) have recently been reviewed in [6]. Simpler optoelectronic device structures essentially based on double barrier resonant tunnelling diode (DBRTDs) have also been used in various applications; these include photodetectors at optical communication wavelengths [7], mid-infrared wavelengths [8] and, optical modulators [9][10][11].

III-V semiconductor optical modulators based on reverse biased *p-n* junction structures and using either the Franz-Keldysh effect in bulk or the quantum confined Stark effect in quantum wells (MQWs) are currently being considered in optical communication systems [1][3]. The application of resonant tunnelling diodes in electroabsorption modulators can be an alternative to conventional *p-i-n* EAM devices. In this paper, we report on high-speed, high on/off ratio, and ultra-low-drive-voltage electroabsorption modulators based on the integration of an InGaAs/AlAs resonant tunnelling diode with an InGaAlAs/InAlAs optical waveguide operating at long wavelengths, the resonant tunnelling diode electroabsorption modulator (RTD-EAM).

## 2. Principle of operation of the RTD-EAM

Essentially, the RTD-EAM is a unipolar device that consists of a DBRTD embedded in a ridge channel optical waveguide configuration lying on top of the substrate. Figure 1 shows a schematic of the RTD-EAM and the Γ-conduction band profile of its layer structure.

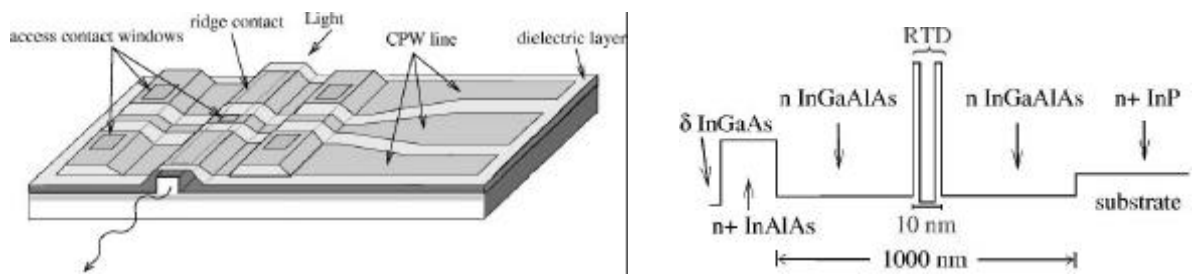

Fig. 1. (a) Schematic diagram of the RTD-EAM. (b) Γ-conduction band profile of the wafer structure.

The presence of the DBRTD within the waveguide core introduces high non-linearities in the current-voltage (I-V) characteristic of the unipolar waveguide. A typical I-V characteristic of a RTD-EAM is shown in Fig. 2 (the physics which gives rise to this type of I-V has been previously explained [6]).

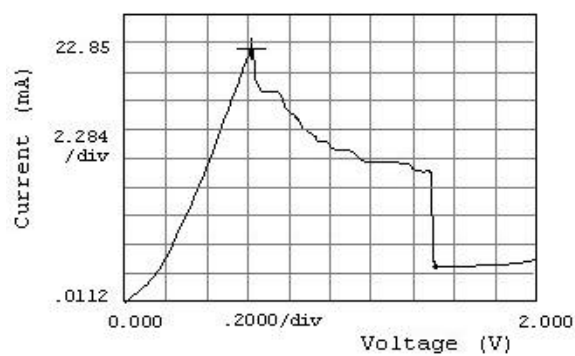

Fig. 2: I-V characteristic of $2 \times 200$ $\mu m^2$ active area RTD-EAM.

The operation of the RTD-EAM is based on a non-uniform electric field distribution across the waveguide induced by the RTD, which becomes strongly dependent on the bias voltage. When the current decreases from the peak to the valley there is an increase of the electric field across the waveguide core. The electric field enhancement in the depleted spacer layer causes the Franz-Keldysh absorption band-edge shift to lower energy that is responsible for the electroabsorption effect.

The physical mechanism by which the current drop is converted into an electric field enhancement is as follows. When the bound state of the RTD quantum well is above or aligned with the emitter conduction band energy minimum (see diagram of Fig. 3) the electron transmission is high and the carriers can easily tunnel through the bound state with little free carrier depletion in the collector region. The applied voltage is dropped mainly across the RTD and the electric field gradient in the collector spacer layer is small because the spacer layer is not strongly depleted. This corresponds to the transmissive state (on-state) of the modulator (during operation the RTD-EAM is dc biased slightly below the peak voltage). Once the applied voltage is increased from the peak to the valley, the RTD bound state is pulled below the emitter conduction band energy minimum, as depicted in Fig. 3, and the electrons can no longer tunnel through using the bound state. The current through the device drops, giving rise to an increase of positive space charge in the collector region; a substantial part of the terminal voltage is now dropped across the collector spacer layer. As a consequence, the magnitude of the electric field in the collector spacer layer increases: this is the non-transmissive state of the modulator. In conclusion, the peak-to-valley current drop produces an increase in the magnitude of the electric field across the waveguide core collector region. This causes the broadening of the waveguide absorption band-edge through the Franz-Keldysh effect to longer wavelengths, which in turn leads to an increase of the optical absorption coefficient of photons possessing energy slightly lower the waveguide core band-edge energy.

In conventional EAM devices the electric field is applied by reverse biasing a *pn* diode that shifts the absorption band-edge of the active region to lower energy. The key difference with the RTD-EAM is that the tunnelling characteristics of the double barrier RTD are employed to switch the electric field across the waveguide collector depleted region. As reported in this paper, a small high frequency ac signal (<<1 V) can induce high speed switching producing substantial modulation of light at photon energy slightly lower than the waveguide band-gap energy. In Fig. 3 we show the energy band diagram in the RTD-EAM at the valley voltage, where the applied voltage is dropped mainly across the depleted region of the waveguide core.

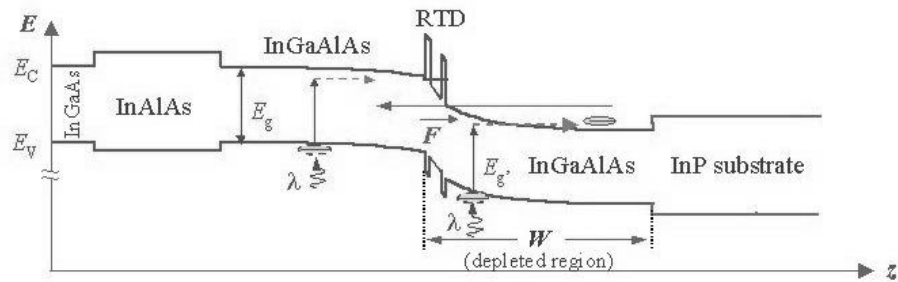

Fig. 3: Schematic diagram of the RTD-EAM energy band diagram at the valley bias voltage.

In essence, the RTD-EAM is a RTD current switch in series with a resistor, except that the speed of response is limited by the electron transit time across the collector depleted spacer layer; if one assumes an electron saturation velocity, $v_{sat}$, of $10^7$ cm/s and a depletion region width $W$~500 nm (see Fig. 3), the electron transit time is 5 ps. The quantitatively analysis of the RTD-EAM can be found in [12].

## 3. RTD-EAM design, fabrication and packaging

The RTD-EAM is a unipolar optical waveguide containing a double barrier RTD, Fig. 1, grown by Molecular Beam Epitaxy on a InP substrate; the RTD is employed to switch the electric field developed across the waveguide collector region as described above and in [12]. The optical waveguide configuration ensures a larger interaction volume between the active region of the device (RTD depletion region) and the waveguide optical mode, thereby ensuring a larger modulation depth for a given applied field. The wavelength of operation is set by the band-gap of the material employed in the active region (waveguide core) of the device. Our initial devices used GaAs in the active region (see [9][10]) and operated at 900 nm, subsequently $In_{0.53}Ga_{0.42}Al_{0.05}As$ was employed to shift the wavelength of operation to 1550 nm (InGaAlAs was used because it is a convenient semiconductor alloy for MBE growth.)

The waveguide core was formed by two moderately doped (Si: $5\times10^{16}$ cm$^{-3}$) $In_{0.53}Ga_{0.42}Al_{0.05}As$ layers 500 nm thick (absorption band edge around 1520 nm and refractive index of 3.56) each side of the DBRTD (2 nm thick AlAs barriers and 6 nm thick $In_{0.53}Ga_{0.47}As$ quantum well). The upper cladding layer of the optical waveguide consisted of a 300 nm $In_{0.52}In_{0.48}As$ layer heavily doped (Si: $2\times10^{18}$ cm$^{-3}$). The contact layer was a $In_{0.53}Ga_{0.47}As$ layer δ-doped for the formation of non-alloyed ohmic contacts.

Ridge waveguides (2 to 6 μm wide) and large-area mesas on each side of the ridges were fabricated by wet etching. Ohmic contacts (100 to 400 μm long) were deposited on top of the ridges and mesas. (The waveguide width and the ohmic contact length define the device active area.) A $SiO_2$ layer was deposited, and access contact windows were etched on the silica over the ridge and the mesa electrodes, allowing contact to be made through high frequency bonding pads (coplanar waveguide transmission line, CPW), Fig. 1(a). After cleaving, the devices were die bonded on packages allowing light to be coupled into the waveguide by a microscope objective/optical fibre end-fire arrangement. The details of the fabrication procedure and device packaging can be found in [13].

## 4. Experimental results and discussion

Here we report on the high frequency optical characterisation of the InGaAlAs/InP RTD-EAM; the electrical characterisation can be found in [12][13]. In the optical experiments described below devices were reverse biased, i.e., electrons flowing towards the substrate.

Low frequency electroabsorption modulation results have been previously reported [11][12]. In summary, devices with 4×200 μm$^2$ active area showing the highest peak-to-valley current (PVCR) and largest peak-to-valley voltage swing $\Delta V_{p-v}$, when dc biased to the optimum operating point, had a maximum modulation depth of 28 dB at around 1565 nm [11]; typical 4×200 μm$^2$ active area devices showed a modulation depth of around 20 dB in the wavelength range 1560-1567 nm, with propagation loss in the transmissive state estimated to be ~5 dB [13].

The RTD-EAM optoelectronic characterisation comprehended the study of the modulator response dependence on the bias point for different injected rf signal voltage amplitude at a given frequency. A signal generator with frequency range 5 kHz to 3 GHz was employed to drive the RTD-EAM, and a continuous light wave with a wavelength of 1550 nm was coupled/decoupled to the waveguide through a single mode fibre (the waveguide was not single mode but it was possible to excite individual modes with a single mode fibre). The modulator optical output was detected using a fibre pigtailed 45 GHz bandwidth Schottky InGaAs photodetector. Figure 4 shows the schematic of the high-frequency characterisation set-up.

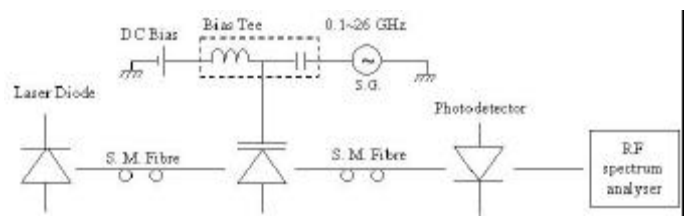

Fig. 4: Schematic diagram of the setup employed for the RTD-EAM optoelectronic characterisation.

Figure 5 shows the 3 GHz rf photodetector output power as a function of the dc bias voltage with the modulator driving rf signal voltage amplitude as a parameter; also represented is dc I-V characteristic of the 4 × 200 μm$^2$ active area device under test.

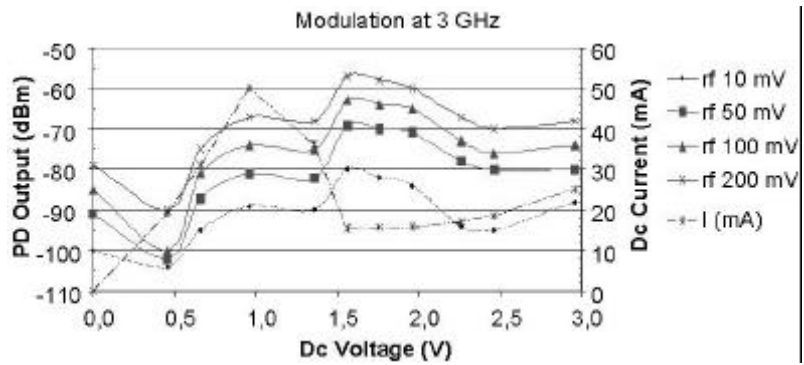

Fig. 5: 3 GHz photodetector output as a function of the dc bias voltage, for several rf voltage amplitudes.

The result shows that the photodetector output power at 3 GHz increases by around 15 dB when the device dc bias point moves from the peak to the valley region of the RTD-EAM I-V characteristic for a driving signal of 3 GHz and amplitude equal to 50 mV. Over certain voltage amplitude around 50 mV, the change in the photodetector output does not vary significantly with the increase of the voltage amplitude. This may indicate that as soon as the electric losses are compensated, the device starts to oscillate at the frequency of the rf driving signal. In collaboration with other groups [14] we have shown previously that with these devices it is possible to achieve a modulation of 5 dB for a voltage change of 1 mV.

Higher frequency optical characterisation of the InGaAlAs RTD-EAMs employed a microwave synthesized signal generator with a maximum output of +20 dBm and an upper frequency limit of 26 GHz. Figure 6 shows the power spectrum of the photodetector output signal at 26 GHz for rf driving power levels of –20 dBm and +7.7 dBm. The RTD-EAM was dc biased in the valley.

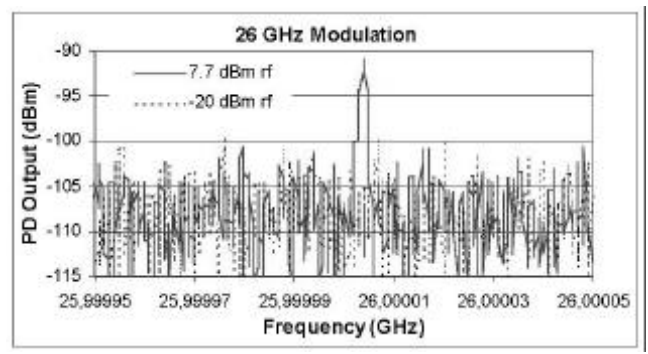

Fig. 6: Spectrum of the 26 GHz photodetected signal at the modulator driving power of –20 dBm and +7.7 dBm.

The results show an increase in the photodetector output signal higher than 10 dB when the device driving rf power rises from –20 dBm to +7.7 dBm at 26 GHz. This is an indication that the device is capable to achieve modulation extinction ratios higher than 10 dB at low power driving levels (<10 mW). We think the results of Fig. 6 are a consequence of on-chip electrical amplification, due to the RTD, which substantially reduces the rf power required from the high frequency signal generator. The device on-chip driving signal amplification can eliminate the need of an external rf amplifier which is required to drive conventional EAM.

The next stage in the development of the InAlGaAs RTD-EAM will be the full high speed electrical and optoelectronic characterisation, after further optimisation of the design of the device and the package.

## 5. Summary

In this paper, we reported results on high frequency characterisation of an electroabsorption waveguide modulator incorporating a resonant tunnelling diode. It was shown that a driving signal as low as +7.7 dBm can produce up to 10 dB modulation depth. This is a consequence of the RTD highly non-linear I-V characteristic and the negative differential resistance.

The intrinsic high speed of the tunnelling process, which has already been demonstrated to operate at over 700 GHz [15] and the capability of combination of optoelectronic and electronic devices makes the RTD-EAM a device with high potential for the next generation of information technology systems, in particular to high-speed communication and signal processing.